# Behaviors of positively charged fine particles in a cross field sheath between magnetized double plasmas

Takuma Gohda and Satoru Iizuka

*Department of Electric Engineering, Graduate School of Engineering, Tohoku University*

*Sendai 980-8579, Japan*

Dependencies of levitation position of positively charged fine-particles on plasma parameters are investigated. The charges on the particles become positive in a cross-field sheath between magnetized double plasmas with different potentials separated vertically by horizontal magnetic field, because ion current flowing from a lower high-potential plasma surpasses electron current coming across the magnetic field from an upper low-potential plasma. From measurement of the resonance frequency of the particles driven by external oscillating electric field, the charge on particle is estimated to be of the order of $10^2$. Variation of particle levitation positions can be explained by the change of the charges.

**I. Introduction**

Dusty plasma containing fine particles of micron size introduced into weakly ionized plasmas reveal various characteristic phenomena concerned with a strongly coupled state of dusty plasmas [1]. Formations of Coulomb crystals, dynamic motion of Coulomb fluid, and various wave phenomena have been reported. By employing a completely dc-discharge plasma, it becomes possible to control the behavior of fine particles more systematically [2]. The effects of vertical magnetic field on the fine-particle behaviors were also investigated and a rotational motion of fine-particle cloud has been observed [3]. Recently, in order to eliminate the effect of the gravity acting on the particles, microgravity experiments have been proposed, and characteristic features of fine particle behaviors have been discovered [4,5]. Under the microgravity condition using a parabolic flight experiment we have observed formations of fine-particle cloud with a spherical void inside and spherical fine-particle cloud with no void inside, respectively [6].

Although almost all previous-experiments are concerned with negatively charged fine particles, we have reported the formation and levitation of positively charged fine particles by using irradiation of slow positive beam ions to the particles [7]. The positive ions with large ionization potential subtract electrons from the neutral particles to create positively charged particles, and as a result the positive beam ions turn to neutral.

**2. Experimental apparatus and method**

Figure 1 shows a schematic of the experimental apparatus. Double plasmas (upper and lower plasmas) are produced independently along the horizontal magnetic field by dc discharges with different anode potentials and separated in the vertical direction by a cross-field sheath. The cathode-anode distance in both plasmas is 4 cm. The anode potential of the upper plasma is grounded, while dc bias voltage $V_A$ can be applied to the anode of the lower plasma. When $V_A > 0$, we can produce electric field directed upward in the cross-field sheath region. In order to avoid a mixing of both plasmas, the strength of the horizontal magnetic field is kept in the range $B = 0.9 – 1.4$ kG. Therefore, we can fix the upper plasma potential almost grounded even when positive $V_A = 10 - 60$ V is applied to the anode of the lower plasma. Details are described in [7].

## 3. Experimental results and discussions

Dependency of floating potential profile $V_F$ in vertical position $z$ on the discharge current of the upper plasma $I_{DU}$ is shown in Fig. 2, where vertical coordinate $z$ is counted from the lower edge of the side wall. Here, the cross-field sheath region is $0 < z < L = 10$ mm. Typically, $B = 1$ kG, $V_A = 30$ V and $P_{Ar} = 20$ Pa. In the lower plasma for $z < 0$ the floating potential $V_F$ become positive and almost uniform in region of $–10$ mm $< z < 0$. On the other hand, in the upper plasma for $z > 10$ mm the potential $V_F$ becomes negative and is almost fixed at ~ -8 V. The potential varies monotonically in vertical direction and the electric field directed upward is formed. In this potential configuration particle cloud is trapped and levitated above the levitation electrode to form Coulomb crystal with spacing of ~ 0.3 mm. Here, the plasma densities in the upper and lower plasmas are $4.2 \times 10^7$/cm$^3$ and $1.4 \times 10^8$/cm$^3$, respectively. The average electron temperature in both plasmas is ~2.6 eV.

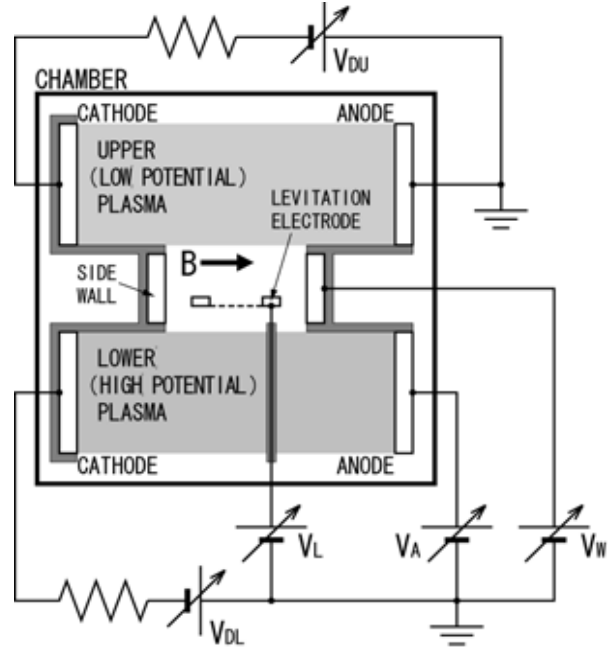

Fig. 1 Experimental apparatus.

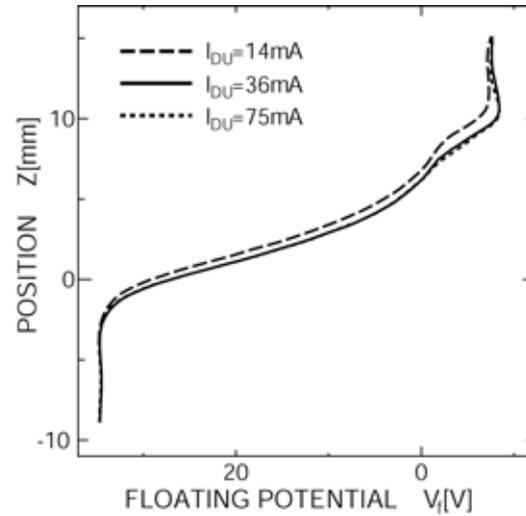

Fig. 2 Variation of floating potential $V_f$ in vertical direction $z$ with discharge current $I_{DU}$ of upper plasma as a parameter.

We measured the variation of particle positions above the levitation electrode as a function of the discharge current $I_{DU}$ of the upper plasma. The position of the particles is shifted downward with an increase in the discharge current $I_{DU}$. The following charging model can explain this behavior. Since the collision mean free path $\lambda_C$ ($\approx 0.4$ mm) of ions with neutral atoms is much shorter than the cross-field sheath width $L$ (= 10 mm), the ions have to suffer many collisions during their motion across the sheath. Therefore, energy distribution function $f_i(W)$ of ions impinging on the fine particles is not a monochromatic, but has an energy spread of the order of $W_C/e \approx (\lambda_C /L)V_A$. Here, $W = (1/2)m_i u_i^2$, $m_i$ is the ion mass, and $u_i$ is the ion velocity. That is, the ions with energy $e\phi_d < W$ can overcome the particle potential $\phi_d$. But, the ions with energy $0 < W < e\phi_d$ are reflected by the positive particle potential $\phi_d$. Since the particles are levitated in the sheath, the condition $I_i = I_e$ for the ion and electron currents flowing into the particles has to be satisfied in the steady state. Therefore, when the electron current $I_e$ coming across the magnetic field from the upper plasma is increased by the increase in the discharge current, $\phi_d$ has to be decreased consequently to enhance $I_i$ to make it equal to the increase in $I_e$. Then, the charge $Q_d$ on the particles decreases, which shifts the particles downward direction for getting a stronger electric field $E$ to satisfy the force balance equation $Q_d E \approx m_d g$ = constant. Here, $g$ and $m_d$ are gravity constant and mass of the particle, respectively.

In order to evaluate the charge on the fine particles we applied small amplitude voltage at frequency $\omega$ to the levitation electrode biased at $V_L$ [8]. By changing the frequency $\omega$ we measure the amplitude of the particles oscillating in vertical direction. The charge can be measured by the resonance frequency and the profile of external electric field. The equation of motion of a particle in the sheath is given by,

$$m_d \frac{d^2 z}{dt^2} = QE_{(z,t)} - m_d g - m_d \beta \frac{dz}{dt},$$

where $z$ is the position in the sheath region, $\beta$ is the damping constant, $Q$ is the charge on the particle. The equilibrium position of the particles in vertical direction is given by $QE_{(z_0)} = m_d g$.

When the external electric field is forced to oscillate periodically with amplitude $\Delta E$ at frequency $\omega$, the electric field is approximately expressed by

$$E_{(z,t)} = E_{(z_0)} \cdot (z - z_0) + \Delta E \sin(\omega t).$$

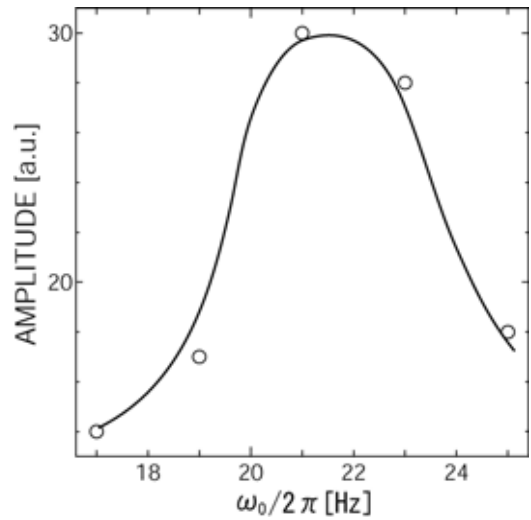

Fig. 3 Particle oscillation amplitude as a function of modulation frequency ω of applied voltage.

From these equations we can solve the vertical motion of particle. The particles move with attenuated oscillation at the frequency $\omega$ with phase delay $\varphi$ around the balance position of $z_0$. The solution has a following form.

$$z_{(t)} = z_0 + A \sin \omega t + B \cos \omega t$$

$$= z_0 + \frac{Q \Delta E}{\sqrt{(m_d \omega^2 + Q \frac{dE_{(z_0)}}{dz})^2 + (m_d \beta \omega)^2}} \sin(\omega t + \varphi),$$

where $\varphi = \tan^{-1}(B/A)$. If we assume the second term in the denominator is negligible, then $\beta \approx 0$. The amplitude of the oscillation will become maximum at the resonance frequency $\omega = \omega_0$ which is obtained from

$$m_d \omega^2 + Q \frac{dE_{(z_0)}}{dz} \approx 0.$$

Then, the charge $Q$ is given by

$$\frac{Q}{e} = -\frac{m_d \omega_0^2}{e\dfrac{dE_{(z_0)}}{dz}}.$$

This means that the charge can be estimated from the resonance frequency $\omega_0$ and the first derivative of the electric field $dE(z_0)/dz$ which can be obtained from Fig. 2. As shown in Fig. 2 the floating potential distribution in the sheath is almost independent of the discharge current $I_{DU}$ of the upper plasma.

Figure 3 shows the amplitude of the particle oscillation as a function of $\omega$. It is shown that the particles resonate around 21 Hz. Dependence of the resonance frequency $\omega_0$ on the discharge current $I_{DU}$ is shown in Fig. 4(a). It is found that the resonance frequency $\omega_0$ increases with a decrease in $I_{DU}$. Figure 4(b) shows the variation of charge $Q/e$ as a function of $I_{DU}$. We find that $Q/e$ decreases with an increase in $I_{DU}$. In the previous report [7] we have shown that the decrease in the discharge current of the upper plasma results in the downward shift of the particle position. Since the electric field in the down side region becomes larger, the charge should be decreased to balance with the gravity. These results are consistent with the result in Fig. 4(b). Therefore, the levitation model of the positive charged particle described here is quite appropriate. Finally, we should note that the diameter of the particles used here is 1.5 μm and the charge $Q/e$ is estimated to be 220 – 170. This amount of charges is also consistent with the charges of $10^3$ – $10^4$ for the negatively charged fine particles of 10 μm in diameter in the conventional plasmas [2].

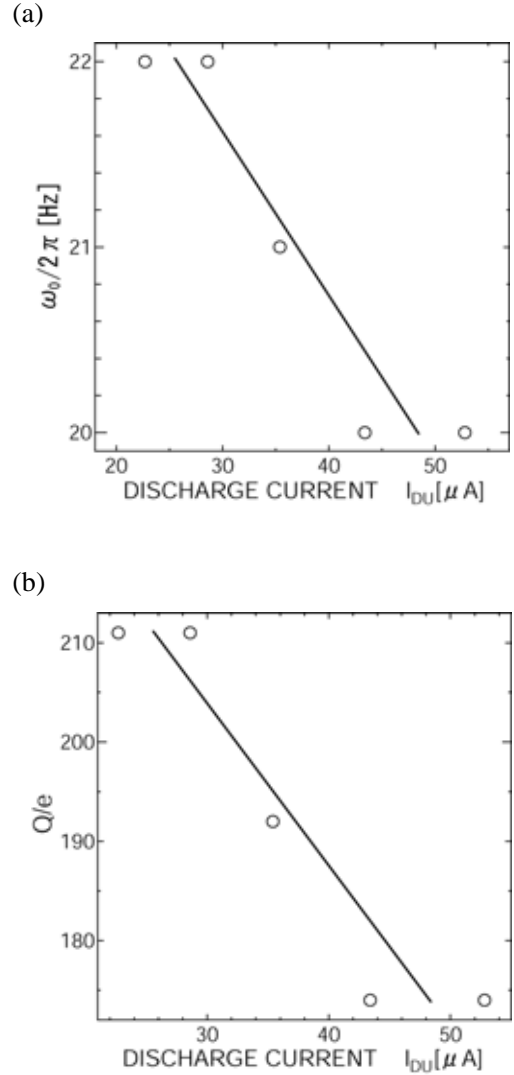

Fig. 4 Variations of (a) resonance frequency $\omega_0$ and (b) charge $Q/e$ as a function of discharge current $I_{DU}$ of the upper plasma.

## 4. Conclusions

We have established a confinement of positively charged particles levitated in the cross-field sheath between magnetized double plasmas. Application of double plasmas and horizontal magnetic field is a key technique for producing upward directed electric field decreasing in the vertical position. We have also measured the positive charges on the particles by applying oscillating potential to the levitation electrode. From the resonance frequency the charge $Q/e$, estimated to be of the order of $10^2$, decreases with an increase in the discharge current of the upper plasma. This dependency is consistent with the particle behavior in vertical direction. The shift of the particle positions can be explained by the change of the charge, caused by the change of the electron flux absorbed by the particles.